 \documentclass[aps,prc,reprint,floatfix,showpacs,showkeys,superscriptaddress]{revtex4-1}
\usepackage{epsf}
\usepackage{graphicx}
\usepackage{wrapfig}
\usepackage{longtable}
\usepackage{array}
\usepackage{color}

\newcommand{\etal}{{\it et al.}}
\def\deg{^\circ}

\begin{document}

\title{Quark Hadron Duality - Recent Jefferson Lab Results}

\author{I.~Niculescu}
\affiliation{James Madison University, Harrisonburg, Virginia 22807}

\date{\today}
\begin{abstract}
Recent Jefferson Lab results on quark-hadron duality are presented.  
\end{abstract}


\maketitle



\section{Introduction}

Inclusive lepton scattering has for many decades been the most important
tool with which to probe the internal quark and gluon (or parton)
structure of nucleons and nuclei.  Structure functions extracted from
inclusive deep-inelastic scattering (DIS) experiments display the central features of quantum chromodynamics (QCD) ---
asymptotic freedom at short distances (via structure function
scaling and its violation) and confinement at large distance scales
(via parton momentum distributions).

Since the late 1960s, DIS experiments have yielded an impressive 
data set that maps nucleon structure functions over several orders
of magnitude in the Bjorken scaling variable, $x$, and the squared
four-momentum transfer, $Q^2$.  These data, supplemented by cross
sections from hadronic collisions and other high-energy processes,
have enabled a detailed picture of the parton distribution functions
(PDFs) of the nucleon which are extracted through global QCD analyses
(see Ref.\ \cite{Jimenez13} and references therein).

At lower energies, where nonperturbative quark--gluon interactions
are important and the inclusive lepton--nucleon cross section is
dominated by nucleon resonances, the structure functions reveal
another intriguing feature of QCD, namely, {\it quark-hadron duality}.
Here, the low energy cross section, when averaged over appropriate
energy intervals, is found to resemble the high energy result,
whose $Q^2$ dependence is described by perturbative QCD.
In this context, quark--hadron duality provides a unique
perspective on the relationship between confinement
and asymptotic freedom, and establishes a critical link between
the perturbative and nonperturbative regimes of QCD.

In the framework of QCD, quark--hadron duality can be formally
interpreted in terms of structure function moments \cite{DeRujula75}.
From the operator product expansion (OPE), the moments can be
expressed as a series in $1/Q^2$, with coefficients given by
matrix elements of local quark--gluon operators of a given twist.
The leading (twist 2) term corresponds to scattering from a single
parton, while higher twist terms correspond to involving multi--quark
and quark--gluon interactions.
Since at low $Q^2$ the resonance region makes a significant
contribution to the structure function moments, one might expect
a strong $Q^2$ dependence of the low-$Q^2$ moments arising from
the higher twist terms of the OPE.
In practice, however, the similarity of the structure function
moments at low $Q^2$ with the moments extracted at high energies
suggests the dominance of the leading twist contribution, with
the higher twist, multi-parton contributions playing a relatively
minor role or that the large higher twists cancel each other.

This non-trivial relationship between the low-energy cross section
and its deep-inelastic counterpart was first observed by Bloom and
Gilman \cite{Bloom70} in the early DIS measurements that were 
instrumental in establishing structure function scaling.
More recently, the availability of extensive, precise structure
function data from Jefferson Lab, over a wide range of kinematics,
has opened up the possibility for in-depth studies of 
quark-hadron duality.  Duality has now been observed
in the proton $F_2$ and $F_L$ structure functions
\ \cite{Niculescu00, Malace09, Melnitchouk05, Monaghan13}, the nuclear
structure function $F_2$\ \cite{niculescu06},
the spin-dependent $g_1$ structure functions of the proton
and $^3$He \cite{Bosted07, Solvignon08}, the
individual helicity-1/2 and 3/2 virtual photoproduction
cross sections for the proton \cite{Malace11}.

To establish the dynamical origin of quark-hadron duality one must also study the {\it neutron}. 
Four-quark higher twists contributions suggest that duality
in the proton could arise from accidental cancellations between
quark charges, which would not occur for the neutron \cite{Brodsky00}.
Unfortunately, the absence of high-density free neutron targets
means that essentially all information on the structure functions
of the neutron has had to be derived from measurements on deuterium.
Typically, the deuterium data are corrected for Fermi smearing and
other nuclear effects \cite{Malace10, Arrington12, Osipenko06,
Hen12, Weinstein11}, which introduces an element of model dependence
into the procedure.  This is particularly problematic in the nucleon
resonance region, where Fermi motion leads to significant
smearing of the resonant structures.
The existence of duality in the neutron $F_2$ structure function
was suggested recently \cite{Malace10} in an analysis which used
an iterative deconvolution method \cite{Kahn08} to extract neutron
resonance spectra from inclusive proton and deuteron $F_2$ data
\cite{Malace09}.  A {\it direct} confirmation of duality in the
neutron, however, was not possible.

%

Recently, a new experimental technique, based on spectator nucleon
tagging \cite{bonus_nim}, has been used to extract the free neutron
$F_2$ structure function \cite{Baillie12}.  By detecting low-momentum
protons at backward angles in electron deuteron scattering,
the BONuS experiment at Jefferson Lab measured $F_2^n$ in both the
resonance and DIS regions, with minimal uncertainty from nuclear
smearing and rescattering corrections \cite{bonus_long}.
In the present paper, we use the BONuS data to quantify for the
first time the degree to which duality holds for the $F_2$ structure
function of the free neutron.  Because the results reported here use
data from an experimentally--isolated neutron target, one can expect
significantly reduced systematic uncertainties compared with those
in the model-dependent analysis of inclusive deuterium data
\cite{Malace10}.

For the theoretical analysis of duality we use the method of
{\it truncated} structure function moments developed by Psaker
{\it et al.}\ \cite{Psaker08}.  Here, the $n$-th truncated moment
of the $F_2$ structure function is defined as
\begin{equation}
M_n(x_{\rm min},x_{\rm max},Q^2)
= \int_{x_{\rm min}}^{x_{\rm max}} dx\, x^{n-2} F_2(x,Q^2),
\end{equation}
\label{eq:moments2}
\noindent where the integration over $x$ is restricted to an interval
between $x_{\rm min}$ and $x_{\rm max}$.
This method avoids extrapolation of the integrand into poorly
mapped kinematic regions, and is particularly suited for the
study of duality where an $x$ region can be defined by a resonance
width around an invariant mass $W^2 = M^2 + Q^2 (1-x)/x$,
where $M$ is the nucleon mass.
As the position of the resonance peak varies with $x$ for
different $Q^2$ values, the values for $x_{\rm min}$ and
$x_{\rm max}$ evolve to correspond to the appropriate invariant
mass squared region.
For the BONuS data, four ranges in $W^2$ were considered,
corresponding to the three prominent resonance regions
($1.3 \leq W^2 \le 1.9$~GeV$^2$ for the first or $\Delta$ resonance region,
 $1.9 \leq W^2 \le 2.5$~GeV$^2$ for the second resonance region, and
 $2.5 \leq W^2 \le 3.1$~GeV$^2$ for the third resonance region),
as well as the combined resonance region
($1.3 \leq W^2 \le 4$~GeV$^2$).
Results on the lowest three non-trivial moments are reported
and compared with recent global PDF parametrizations, as well as
with previous model-dependent data analyses.

In Sec.\ \ref{sec:experiment}, we review the BONuS experiment and the results for the neutron $F_2$ structure function. The analysis of the truncated moments is discussed in Sec.\ \ref{sec:truncated}, together with the implications for duality and its violation. Finally, in Sec.~\ref{sec:conclusion} we summarize our results and discuss their wider implications.

\section{The BONuS Experiment}
\label{sec:experiment}

The results reported here rely on a novel experimental technique aimed at eliminating or substantially
reducing the theoretical uncertainties involved in extracting neutron data from nuclear targets. The BONuS (Barely Off--shell Nucleon Structure) experiment\ \cite{bonus_nim, Baillie12,bonus_long}
used a Radial Time Projection Chamber (RTPC) to detect backward--moving, low momentum spectator protons 
produced in electron--deuterium scattering in conjunction with electrons detected using CLAS\ \cite{clas_nim} in Hall B at Jefferson Lab.
By tagging low momentum backward moving spectator protons one minimizes final state interactions\ \cite{cdg04, Cosyn11, Cosyn14} and ensures that the neutron is just barely
off--shell\ \cite{Baillie12}. Additionally, Fermi smearing effects are essentially eliminated.

The BONuS experiment ran in 2005 and acquired electron--deuteron scattering data at two electron
beam energies: 4.223, and 5.262~GeV. 
The RTPC consists of three layers of gas electron multipliers surrounding a thin, pressurized gas deuterium target which can detect spectator protons with momenta as low as 70~MeV/c. 
The scattered electron was detected by CLAS.
The experiment and data analysis are described in detail in\ \cite{bonus_long}. Ratios of neutron to proton $F_2$ structure functions and the neutron $F_2$ structure function by itself were extracted over a wide kinematic range and for proton spectator momenta between 70 and 100 MeV/c. The total systematic uncertainty in the neutron structure function extracted is 8.7\%\ \cite{bonus_long}. Additionally, there is an overall 10\% scale uncertainty due to cross normalization of the BONuS data to existing $F_2^n/F_2^d$ parameterizations.

\begin{figure}
\begin{center}
\includegraphics[scale=0.4]{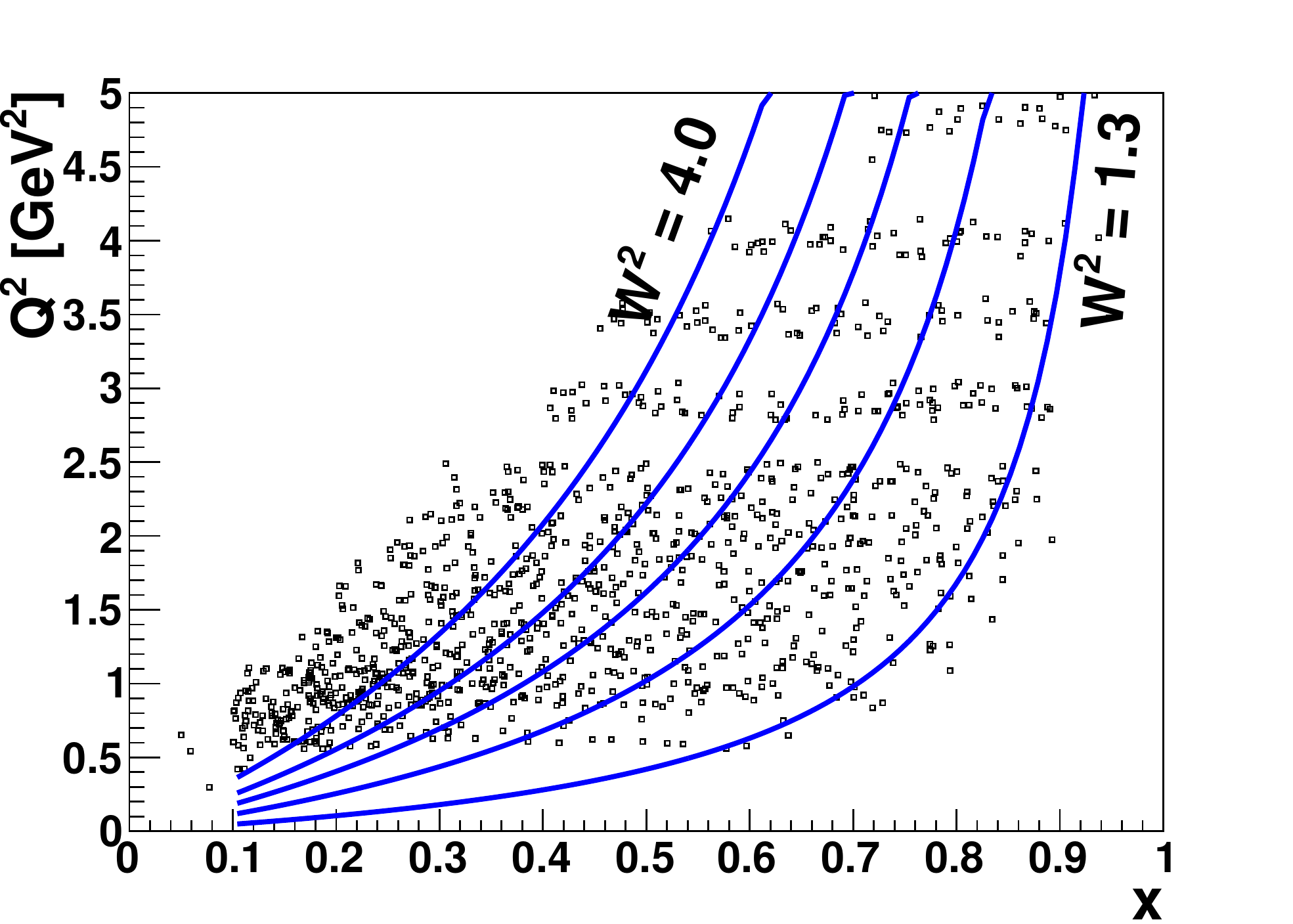}
\end{center}
\caption{Kinematic coverage of the BONuS data. The lines represent the $W^2$ thresholds
for the four resonance mass regions mentioned in the text.}
\noindent
\label{fig:bonus_kinematics}
\end{figure}
\begin{figure}
\begin{center}
\includegraphics[scale=0.38]{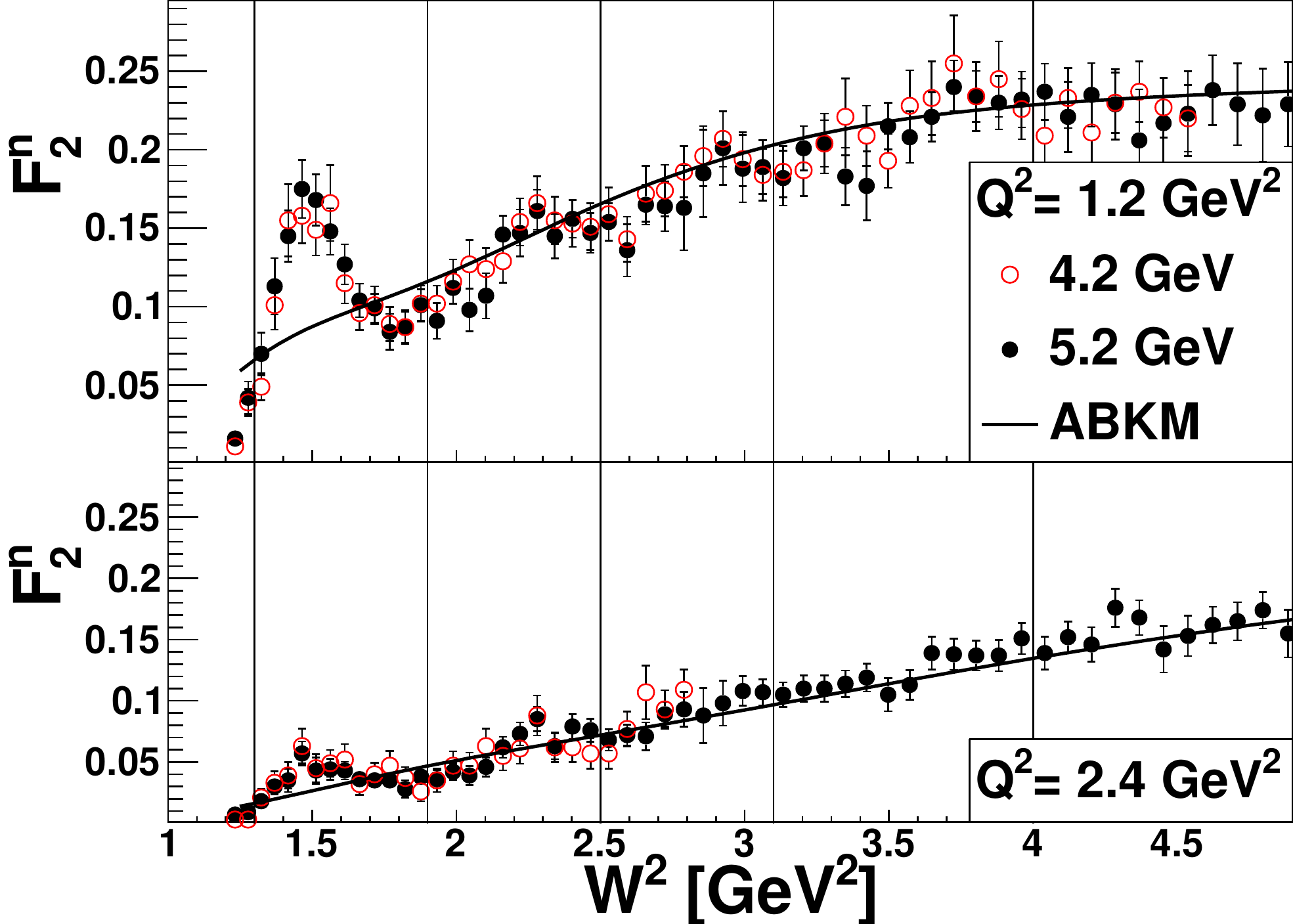}
\end{center}
\caption{Typical neutron structure function data from the BONuS experiment at $Q^2=1.2$~GeV$^2$ (top panel) and $Q^2=2.4$~GeV$^2$ (bottom panel). The open circles represent data obtained with a beam energy of $E=4.2$~GeV, while the closed circles were obtained with $E=5.2$~GeV beam energy. The curves shown are the ABKM DIS parameterizations\ \protect\cite{ABKM} with (solid line) and without (dashed line) higher twist effects and target mass corrections.}
\noindent
\label{fig:spectra}
\end{figure}

The kinematic coverage shown in Figure\ \ref{fig:bonus_kinematics} (4.223 and 5.262~GeV results combined)
 extends from the quasielastic peak to the deep inelastic region
corresponding to final state invariant masses of $W^2 \geq 5$~GeV$^2$. The curves shown
represent the $W^2$ thresholds for the resonance mass regions mentioned above. Typical $F_2^n$ results for $Q^2=1.2$ and $2.4$~GeV$^2$ are shown in Figure\ \ref{fig:spectra}.

The open/closed symbols correspond to data from 4.2 and 5.2~GeV electron beam energies, respectively. Predictions, with and without higher twist effects,
of the ABKM QCD fit\ \cite{ABKM} are also shown. Qualitatively, one can see evidence here for quark--hadron duality in that the curves generally go through the resonance data.

The data used in this analysis 
had spectator angles with respect to the momentum transfer greater than 100$\deg$ and momenta between  70 and 100~MeV/c.

\section{Truncated Moments and Local Quark--Hadron Duality}
\label{sec:truncated}

\begin{figure}
\begin{center}
\includegraphics[scale=0.42]{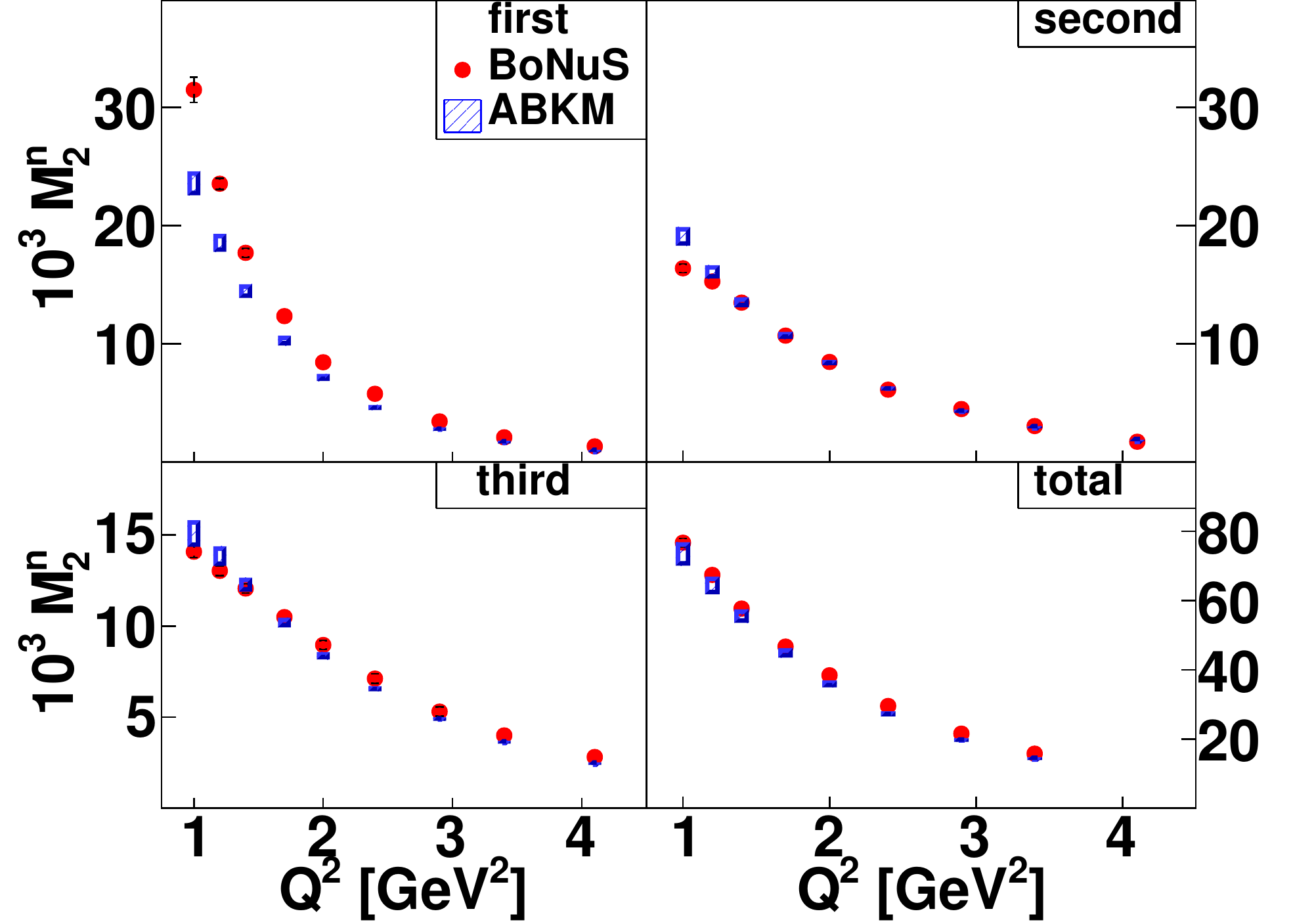}
\end{center}
\caption{The second neutron truncated moments $M_2$, as a function of $Q^2$. The closed circles represent
the moments obtained
from the BONuS data, while the blue rectangles are the moments obtained from the ABKM parameterization
\ \protect\cite{ABKM} which includes target mass and higher twist corrections.}
\noindent
\label{fig:moments}
\end{figure}

Since $Q^2$, $x$, and $W^2$ are not independent of each other, a range in $W^2$ at fixed $Q^2$
implies a corresponding range in $x$. This allows for a straightforward integration of the experimental $F_2^n$ structure function data to obtain truncated moments. In order to minimize model dependence, the integrals were evaluated based solely on the
experimentally measured points without using any inter- or extrapolating function. The second ($n=2$) truncated moments, $M_2$, obtained from these data are shown in Fig.\ \ref{fig:moments} as a function of $Q^2$, and are listed in Table\ \ref{tab:moments}. The uncertainties
quoted take into account the experimental statistical and systematic uncertainties but do not the include the 10\% scale uncertainty due to cross normalization of the BONuS data. The corresponding higher order truncated moments ($n=4$ and $n=6$) are provided in Tables\ \ref{tab:moments4} and\ \ref{tab:moments6}, respectively.

\begin{table}
\begin {tabular}{|l|l|l|l|l|} \hline
& \multicolumn{4}{|c|} {1000*$M_2$}   \\ \hline
$Q^2$ (GeV$^2$) & first & second & third & whole \\ \hline
1.00 & 31.5 $\pm$ 1.1 & 16.4 $\pm$ 0.4 & 14.1 $\pm$ 0.3 & 76.7 $\pm$ 1.2 \\
1.20 & 23.5 $\pm$ 0.5 & 15.3 $\pm$ 0.3 & 13.0 $\pm$ 0.3 & 67.4 $\pm$ 0.6 \\
1.40 & 17.7 $\pm$ 0.4 & 13.5 $\pm$ 0.2 & 12.1 $\pm$ 0.3 & 57.7 $\pm$ 0.5 \\
1.70 & 12.3 $\pm$ 0.3 & 10.7 $\pm$ 0.2 & 10.5 $\pm$ 0.2 & 46.7 $\pm$ 0.5 \\
2.00 & 8.4 $\pm$ 0.2 & 8.5 $\pm$ 0.2 & 9.0 $\pm$ 0.2 & 38.4 $\pm$ 0.4 \\
2.40 & 5.8 $\pm$ 0.2 & 6.1 $\pm$ 0.1 & 7.1 $\pm$ 0.3 & 29.5 $\pm$ 0.4 \\
2.90 & 3.4 $\pm$ 0.1 & 4.5 $\pm$ 0.1 & 5.3 $\pm$ 0.3 & 21.5 $\pm$ 0.4 \\
3.40 & 2.1 $\pm$ 0.1 & 3.1 $\pm$ 0.1 & 4.0 $\pm$ 0.2 & 15.8 $\pm$ 0.3 \\
4.10 & 1.3 $\pm$ 0.1 & 1.7 $\pm$ 0.1 & 2.8 $\pm$ 0.1 &  N/A  \\ \hline \hline
\end{tabular}
\caption{Second order truncated moments of the neutron $F_2$ structure function based on the BONuS data for the three resonance regions studied, as well as for the whole region $1.3\leq W^2\leq 4$~GeV$^2$. For $Q^2=4.10$~GeV$^2$ the moment for the whole region was not computed due to the lack of data in the highest $W^2$ region.}
\label{tab:moments}
 \end{table}

\begin{table}
\resizebox{8.5 cm}{!}{
\begin {tabular}{|l|l|l|l|l|} \hline
& \multicolumn{4}{|c|} {1000*$M_4$}   \\ \hline
$Q^2$ (GeV$^2$)& first & second & third & whole \\ \hline
1.00 & 11.58 $\pm$ 0.43 & 3.09 $\pm$ 0.08 & 1.69 $\pm$ 0.04 & 17.49 $\pm$ 0.44 \\
1.20 & 9.80 $\pm$ 0.21 & 3.51 $\pm$ 0.06 & 1.95 $\pm$ 0.04 & 16.78 $\pm$ 0.22 \\
1.40 & 8.11 $\pm$ 0.17 & 3.60 $\pm$ 0.06 & 2.17 $\pm$ 0.04 & 15.61 $\pm$ 0.19 \\
1.70 & 6.27 $\pm$ 0.14 & 3.40 $\pm$ 0.06 & 2.33 $\pm$ 0.05 & 14.01 $\pm$ 0.17 \\
2.00 & 4.67 $\pm$ 0.14 & 3.08 $\pm$ 0.06 & 2.36 $\pm$ 0.06 & 12.45 $\pm$ 0.17 \\
2.40 & 3.48 $\pm$ 0.11 & 2.54 $\pm$ 0.06 & 2.20 $\pm$ 0.08 & 10.59 $\pm$ 0.15 \\
2.90 & 2.22 $\pm$ 0.10 & 2.11 $\pm$ 0.07 & 1.93 $\pm$ 0.09 & 8.52 $\pm$ 0.16 \\
3.40 & 1.44 $\pm$ 0.09 & 1.58 $\pm$ 0.07 & 1.64 $\pm$ 0.08 & 6.72 $\pm$ 0.15 \\
4.10 & 0.95 $\pm$ 0.08 & 0.98 $\pm$ 0.07 & 1.29 $\pm$ 0.06 & N/A  \\ \hline \hline
\end{tabular}}
\caption{Fourth order truncated moments of the neutron $F_2$ structure function based on the BONuS data for the three resonance regions studied, as well as for the whole region $1.3\leq W^2\leq 4$~GeV$^2$. For $Q^2=4.10$~GeV$^2$ the moment for the whole region was not computed due to the lack of data in the highest $W^2$ region.}
\label{tab:moments4}
 \end{table}

\begin{table}
\resizebox{8.5 cm}{!}{
\begin {tabular}{|l|l|l|l|l|} \hline
& \multicolumn{4}{|c|} {1000*$M_6$}   \\ \hline
$Q^2$ (GeV$^2$)& first & second & third & whole \\ \hline
1.00 & 4.39 $\pm$ 0.18 & 0.60 $\pm$ 0.01 & 0.20 $\pm$ 0.01 & 5.28 $\pm$ 0.18 \\
1.20 & 4.19 $\pm$ 0.10 & 0.82 $\pm$ 0.01 & 0.30 $\pm$ 0.01 & 5.45 $\pm$ 0.10 \\
1.40 & 3.79 $\pm$ 0.09 & 0.98 $\pm$ 0.02 & 0.39 $\pm$ 0.01 & 5.38 $\pm$ 0.09 \\
1.70 & 3.24 $\pm$ 0.08 & 1.09 $\pm$ 0.02 & 0.52 $\pm$ 0.01 & 5.17 $\pm$ 0.08 \\
2.00 & 2.62 $\pm$ 0.08 & 1.13 $\pm$ 0.02 & 0.62 $\pm$ 0.02 & 4.82 $\pm$ 0.09 \\
2.40 & 2.12 $\pm$ 0.07 & 1.06 $\pm$ 0.02 & 0.68 $\pm$ 0.02 & 4.41 $\pm$ 0.08 \\
2.90 & 1.45 $\pm$ 0.07 & 1.00 $\pm$ 0.03 & 0.71 $\pm$ 0.03 & 3.77 $\pm$ 0.08 \\
3.40 & 0.99 $\pm$ 0.07 & 0.82 $\pm$ 0.04 & 0.67 $\pm$ 0.03 & 3.14 $\pm$ 0.09 \\
4.10 & 0.68 $\pm$ 0.06 & 0.56 $\pm$ 0.04 & 0.60 $\pm$ 0.03 &  N/A  \\ \hline \hline
\end{tabular}}
\caption{Sixth order truncated moments of the neutron $F_2$ structure function based on the BONuS data for the three resonance regions studied, as well as for the whole region $1.3\leq W^2\leq 4$~GeV$^2$. For $Q^2=4.10$~GeV$^2$ the moment for the whole region was not computed due to the lack of data in the highest $W^2$ region.}
\label{tab:moments6}
 \end{table}

\begin{figure}
\begin{center}
\includegraphics[scale=0.38]{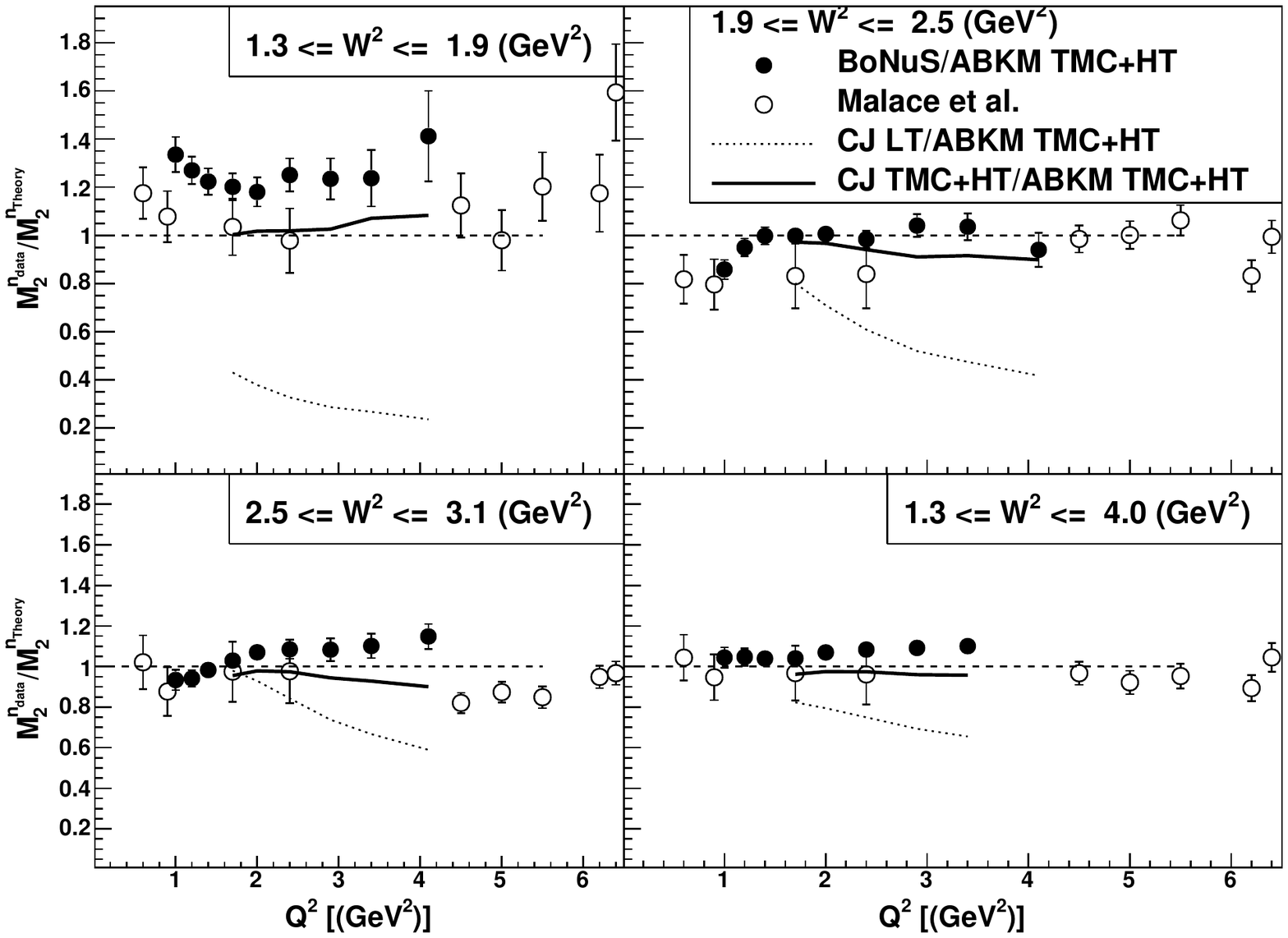}
\end{center}
\caption{The ratio between the $M_2$ truncated moments for the neutron structure function $F_2$ obtained from the BONuS data and the same quantity obtained from the ABKM parameterization\ \protect\cite{ABKM} (closed circles). The ABKM calculation includes both higher twists and target mass corrections. The open circles represent the model--dependent results obtained by\ \protect\cite{Malace10}. The lines represent comparisons of the ABKM with the CJ12 parameterizations\protect\cite{CJ12}: thick solid line for the case when TMC and HT effects are included in both models, thin dashed line when the CJ model is evaluated only for leading twist.}
\noindent
\label{fig:data_theory_m2}
\end{figure}

In order to study local duality, we formed the ratio of the truncated moments of
$F_2^n$ in the resonance region obtained from the BONuS data to the neutron moments
calculated over the same $x$ range based on the ABKM QCD fit\ \cite{ABKM}. These
truncated moments are shown as a function of $Q^2$ in Fig.\ \ref{fig:data_theory_m2} (closed circles)
for $M_2$. For the $M_2$ the earlier, model--dependent results from\ \cite{Malace10} are also shown (open circles). The four panels correspond to the four invariant mass regions previously defined. The computed ABKM moments include higher twist effects and target mass corrections, which allows for a direct comparison of the present results with the studies from Ref.\ \cite{Malace10}. The figure also presents the ratio of truncated $F_2$ neutron moments computed using the CTEQ--Jlab (CJ12) parameterization\ \cite{CJ12} with respect to the same ABKM moments. The explicit inclusion of higher twist terms in these parameterization allows for extending the region of validity into the large--$x$ region. The interplay between higher twist and target mass corrections and their influence on the PDFs at large $x$ is studied in depth in Ref.\ \cite{CJ10}. In addition, the CJ12 parameterization includes nuclear effects in part to improve the neutron, and hence $d$-quark PDF. 

To study the possible influence of target mass corrections and higher twists the CJ12 model was used to obtain both the leading twist (LT, thin dashed line) and the target mass and higher twist corrected (TMC+HT, thick solid line) predictions. The differences between
the two models including higher twist and target mass corrections presented
are within the uncertainty of the data. In the resonance region, target mass corrections and higher twist effects are expected to be sizable and so it may not be surprising that agreement of the data moments with leading twist parameterizations is improved by the inclusion of HT and TMC effects. However, the ABKM and CJ12 parameterizations extracted TMC and HT corrections from deep inelastic scattering data at higher $Q^2$. Therefore confirmation of duality indicates that these contributions are similar in the resonance region and in the DIS.  An alternative possibility is that the PDF fits have a very large uncertainty at high $x$ and perhaps need to increase the $d(x)$ values to better describe the data.

In Fig.\ \ref{fig:data_theory_m2} the first $W$ range corresponding to the $\Delta$ resonance displays the same $Q^2$ behavior, but is 20--40\% larger than the model expectation. This underscores the likelihood that the PDF--based fits underestimate the large $x$ regime, to which this $W$ region corresponds.

Although the BONuS results reported here are in good agreement
with the earlier data extracted from deuteron using modeling (open circles)\ \cite{Malace10}, and both are consistent with unity  in the second and third resonance region,
the BONuS data are consistently higher in the first resonance region. Although all local duality studies on the neutron are sensitive to binding effects, Fermi motion, and final state interactions, obtaining neutron structure functions by subtracting (smeared) proton from deuteron data is especially sensitive to the nuclear models employed at the largest $x$ (i.e. smallest $W$) values. Therefore, this discrepancy could be due to either a difference in the free neutron/proton $\Delta$ resonance excitation or a manifestation of nuclear theory uncertainties. Interestingly, the BONuS/CJ12 ratios assuming no higher twist, would fall consistently below unity, as evidenced by the dashed line.

Moments in the resonance region display the same $Q^2$ dependence as the PDF predictions, i.e. the ratio of the two shown in Fig.\ \ref{fig:data_theory_m2} is independent of $Q^2$.  
This is a remarkable confirmation of one aspect of duality, {\it i.e.} that the large $x$ and lower $Q^2$, region can be described by the same evolution equations -- displaying the same dynamics -- as the higher $Q^2$, larger $W$, scattering regime of single quark deep inelastic scattering. 
This is true even for the very largest $x$, lowest $W$, area. Here, the uncertainties on the PDFs are large and could easily account for the $\sim 20$\% strength difference, but not the $Q^2$--dependence. 


\begin{figure}
\begin{center}
\includegraphics[scale=0.42]{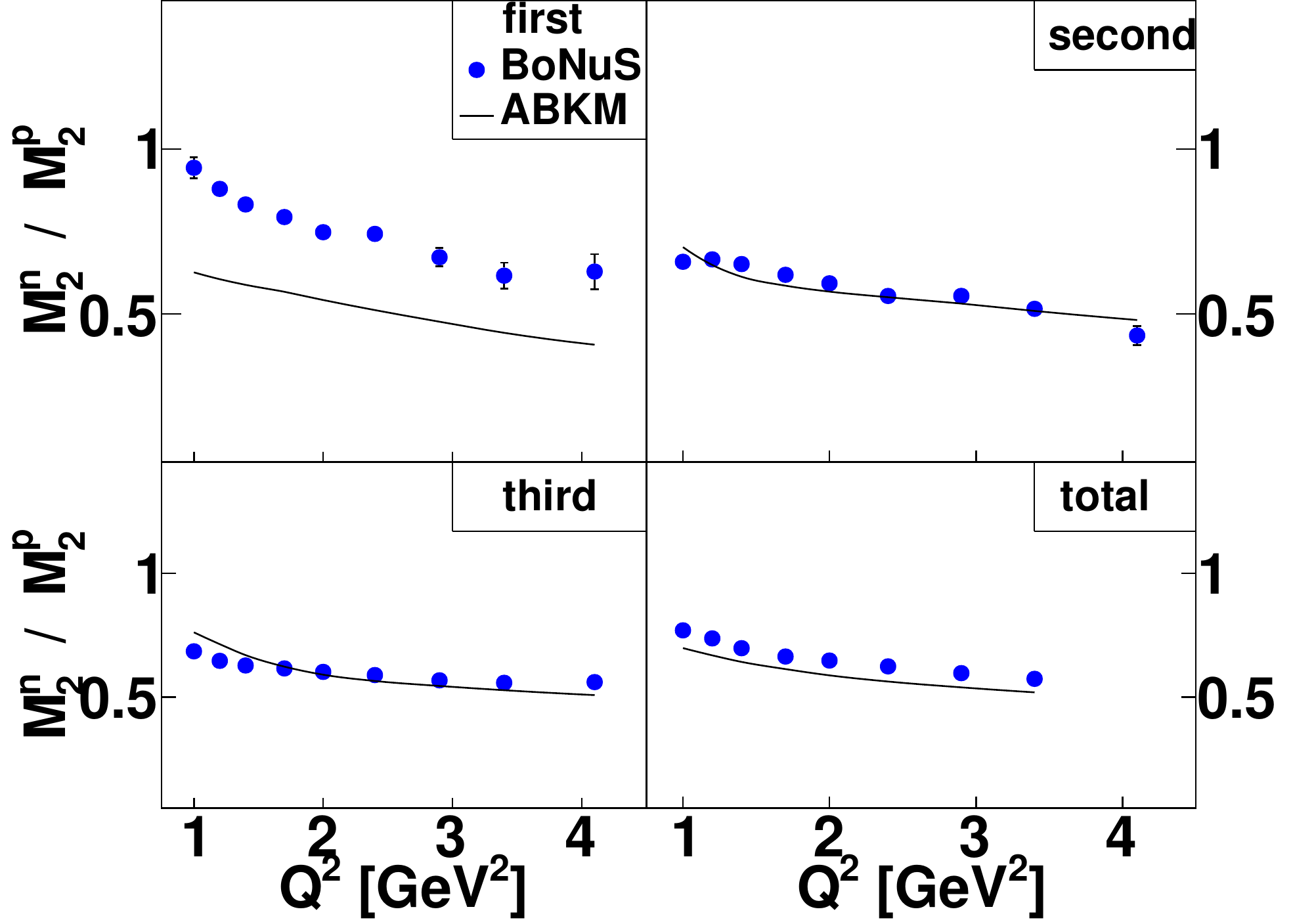}
\end{center}
\caption{The ratio of neutron to proton $F_2$ truncated moments, $M_2^n/M_2^p$ for the four regions studied.
The closed circles represent
the BONuS results, while the solid line corresponds to the ABKM QCD fit prediction including HT and TMCs.}
\noindent
\label{fig:data_f2nf2p}
\end{figure}

Local quark--hadron duality in the proton $F_2$ structure function has been studied
extensively\ \cite{Malace09, Niculescu00}, showing possible violations mainly in the
first resonance region.
To compare the proton with the neutron results, truncated $F_2$ moments were constructed for the
proton using the ABKM model to calculate the deep inelastic moment and a global fit\ \cite{Christy2010} for the resonance region. Fig.\ \ref{fig:data_f2nf2p} shows
the ratio of the neutron--to--proton truncated moments (closed circles) as a function of $Q^2$ for the
four resonance regions. The line shown is the ratio of neutron and proton
truncated moments calculated using the ABKM (including HT and TMC) model. The model agrees well with
the data in the second and third resonance regions but substantially underestimates the data
in the first resonance region. This can be indicative of a fundamental difference between the proton
and the neutron at the very largest $x$ values corresponding to the $\Delta$ resonance region.
Alternatively, this can be the result of the fact that QCD fits for either proton or neutron are not well constrained at large $x$, which is the kinematic range of the first resonance region.



 In conventional parameterizations of parton distribution
functions the $d/u$ ratio is often assumed to go either to zero or infinity as $x\to 1$, depending on the parameterization of the $d$ PDF\ \cite{CJ11}. At $x$ above 0.8 the uncertainties on the PDFs, especially 
$d$, are considerable, due to deuteron nuclear model corrections and to the lack of large $x$ data.
 The ratio $d/u$, and correspondingly $F_2^n/F_2^p$, at large $x$ is however of fundamental interest\ \cite{Isgur99,Brodsky95,Meln96} and there are numerous QCD--based theoretical predictions for this quantity. All require $d(x)< u(x)$ and hence $F_2^n < F_2^p$ for a given $x$ and $Q^2$. This is in contrast to naive predictions for the $\Delta$ resonance. For the latter, given that the proton and
neutron transitions to the $\Delta$ are isovector, the resonant contributions should be
identical, which would lead to an $F_2^n/F_2^p$ ratio close to unity. It is perhaps not
surprising that the resonance behavior dominates in this $W$ region because the $\Delta$ region contains  only one resonance with less continuum background than the other regions.



Similar tension between the resonance data and DIS/ pQCD-derived moments could exist
in the second and third resonance regions as well. Assuming a dominance of magnetic couplings, 
as some quark models do\ \cite{Close01, Close03, Close09},  the proton resonance data should
overestimate the DIS prediction in these regions due to the odd--parity resonances, such as the
prominent spin$-1/2$ octet, resulting in a measured neutron--to--proton truncated moment ratio
that should fall below the corresponding DIS curve. As shown in Fig.\ \ref{fig:data_f2nf2p} the
results presented here do not support this argument, and duality is apparently the predominant effect for regions above the $\Delta$.


\section{Conclusion}
\label{sec:conclusion}

In conclusion, this paper investigates local quark--hadron duality in the neutron
structure function based on data obtained by the BONuS experiment at Jefferson Lab, which used 
a novel experimental technique to tag the spectator proton in a deuterium target and thereby create an effective neutron target. This technique provides smaller systematic uncertainties than
earlier studies that relied on the subtraction of smeared hydrogen data and nuclear modeling from deuterium.
Truncated $F_2$ structure function moments were compared to PDF fits based on pQCD and largely
deep inelastic scattering data, as well as to similar truncated moments obtained for the proton.
The results indicate that quark--hadron duality holds dynamically everywhere, as well as in sum
 for the second and third resonance
regions. This data provides further confirmation that duality, still not well understood, appears to be a fundamental aspect of nucleon structure.



\end{document}